\documentclass[dvipdfmx]{article}
\newtheorem{theorem}{Theorem}%
\newtheorem{corollary}{Corollary}%
\newcommand{\qed}{\hbox{\rule[-2pt]{3pt}{6pt}}}

\begin{document}

\begin{center}

{\large 
Onsager algebra and \\
algebraic generalization 
of Jordan-Wigner transformation}
\vspace{0.6cm}

Kazuhiko Minami
\vspace{0.6cm}

Graduate School of Mathematics, Nagoya University, \\
Furo-cho, Chikusa-ku, Nagoya, Aichi, 464-8602, JAPAN.

\end{center}

\begin{abstract}
Recently, an algebraic generalization of the Jordan-Wigner transformation 
was introduced and applied to one- and two-dimensional systems. 
This transformation is composed of the interactions $\eta_{i}$ 
that appear in the Hamiltonian ${\cal H}$ 
as ${\cal H}=\sum_{i=1}^{N}J_{i}\eta_{i}$, 
where $J_{i}$ are coupling constants. 
In this short note, 
it is derived that 
operators that are composed of $\eta_{i}$, 
or its $n$-state clock generalizations, 
satisfy the Dolan-Grady condition 
and hence obey the Onsager algebra 
which was introduced in the original solution of the rectangular Ising model 
and appears in some integrable models. 
\end{abstract}

\noindent
Keywords: lattice spin system, Jordan-Wigner transformation, Onsager algebra, exact solution, integrable system

\noindent
e-mail:minami@math.nagoya-u.ac.jp
\\
tel.+81-52-789-5578, fax:+81-52-789-2829

\section{Introduction}

When Onsager solved in \cite{44Onsager} the square-lattice Ising model, 
the basic structure of his derivation is an algebraic structure 
which is now called the Onsager algebra. 
He showed that 
the representation of the original Hamiltonian can be reduced into 
direct products of two-dimensional representations. 
This decomposition intrinsically suggests the structure of the free fermion system, 
although he did not at all used the word 'fermion' in his paper. 
Later, Kaufman \cite{49Kaufman} rederived the partition function of the model, 
with the use  of generators of the Clifford algebra, 
and later Schultz, Mattis and Lieb \cite{64SchultzMattisLieb} rederived the free energy 
through a direct transformation to the free fermion system. 
The transformations in both \cite{49Kaufman} and \cite{64SchultzMattisLieb} 
were the Jordan-Wigner transformation.

In 1982, Dolan and Grady \cite{82DolanGrady} constructed 
an infinite number of conserved charges 
for a self-dual Hamiltonian that satisfy a condition, 
which is now called the Dolan-Grady condition. 
Later, von Gehlen and Rittenberg \cite{85GehlenRittenberg}
introduced a $n$-state chiral Potts model 
with specific coupling constants. 
They derived that 
this model is integrable in the sense 
that it satisfies the Dolan-Grady condition, 
and hence there exist an infinite number of conserved charges, 
and also they numerically showed that this model 
exhibits an Ising-like spectrum. 
This model was also investigated in \cite{89AlbertiniMcCoyPerkTang} 
and called "superintegrable", 
because it obeys the Onsager algebra, 
in addition to showing the structure of commuting transfer matrices.

It was pointed out in \cite{89Perk} that 
specific operators appearing in \cite{82DolanGrady} 
satisfy the defining relations of the Onsager algebra. 
Davies derived \cite{91Davies} that 
a pair of operators recursively generate the Onsager algebra 
provided that they satisfy 
two symmetric Dolan-Grady relations; 
self-duality of the Hamiltonian is not needed in this argument. 

Let us summarize the following progresses. 
The irreducible representations 
of finite-dimensional Onsager algebra were obtained,  
and the general form of the eigenvalues of the associated Hamiltonians $A_{0}+kA_{1}$, where $k$ is the coupling constant, 
were determined in \cite{90Davies}, 
and  subsequently in \cite{91Roan}. 
Lie algebraic structure of the Onsager algebra was investigated 
in \cite{00DateRoan2}. 
The Onsager algebra is a subalgebra of $\widehat{sl_{2}}$ 
\cite{91Davies}\cite{90Davies}\cite{20Stokman}, 
related with a class of Yang-Baxter algebras \cite{18BaseilhacBelliardCrampe}. 
Integrable lattice models 
were derived based on the Onsager algebra 
and an extension of the Onsager algebra was also considered 
in \cite{90AhnShigemoto}. 
Higher rank generalizations of the Dolan-Grady relations and Onsager algebras 
have been also investigated starting from \cite{95UglovIvanov}\cite{04DateUsami} 
and more recently in \cite{20Stokman}\cite{18BaseilhacBelliardCrampe}. 
The completely inhomogeneous transverse Ising chain 
was considered in \cite{95UglovIvanov}. 
A contracted case of the Dolan-Grady condition 
and related spin models were considered in 
\cite{02KlishevichPlyushchay}\cite{03KlishevichPlyushchay}.
A q-deformed analogue of Onsager's symmetry 
was introduced in \cite{05BaseilhacKoizumi}. 
It was shown in \cite{05BaseilhacKoizumi2}
that the homogeneous XXZ open spin chain with integrable boundary conditions 
can be built from the generators of the $q$-Onsager algebra, 
and eigenstates were investigated 
in the thermodynamic limit in \cite{06Baseilhac},
its transfer matrix was diagonalized in \cite{07BaseilhacKoizumi},  
for a review see also \cite{07Baseilhac}.
The Onsager symmetry appears in a kind of $n$-state clock chains 
\cite{19VernierOBrienFendley} 
whose $Z_n$ symmetry is enhanced to U(1). 
It was derived that the Hamiltonian in \cite{19VernierOBrienFendley} with additional terms 
exhibits \cite{20ShibataYoshiokaKatsura} the scar states. 
Motivated by the results in \cite{19VernierOBrienFendley}, 
the spin-1/2 XXZ chain at root of unity was investigated 
\cite{21MiaoLamersPasquier}, 
and the existence of the Onsager symmetry 
at root of unity was conjectured 
\cite{21Miao}. 
Dynamics of models where the Hamiltonian is an element of the Onsager algebra 
were also investigated in \cite{21Lychkovskiy}. 
\\

There exists another progress on solvable models. 
Recently, an algebraic generalization of the Jordan-Wigner transformation 
was introduced \cite{16Minami}. 
This formula can be summarized as follows: 
Consider a series of operators $\{\eta_j\}$ $(j=1, 2, \ldots, M)$  
that satisfy the following commutation relations 
\begin{eqnarray}
\eta_{j}\eta_{j+1}=-\eta_{j+1}\eta_{j},
\hspace{0.6cm}
\eta_{j}\eta_{k}=\eta_{k}\eta_{j}
\hspace{0.3cm}
(|j-k|>1),
\hspace{0.6cm}
\eta_{j}^{2}=1.
\label{cond2}
\end{eqnarray}
Then, the Hamiltonian 
$\displaystyle -\beta{\cal H}=\sum_{j=1}^{M}K_j \eta_j$
is mapped to the free fermion system by the following transformation: 
\begin{eqnarray}
\varphi_j
=
\frac{1}{\sqrt{2}}
e^{i\frac{\pi}{2}(j-1)}
\eta_0
\eta_1
\eta_2
\cdots
\eta_j
\hspace{0.8cm}
(0\leq j\leq M-1), 
\label{transmain}
\end{eqnarray}
where $\eta_0$ is an initial operator satisfying 
$\eta_0^2=-1$,  $\eta_0\eta_1=-\eta_1\eta_0$,  
and $\eta_0\eta_k=\eta_k\eta_0$ $(2\leq k\leq M)$. 
Then we obtain $(-2i)\varphi_{j}\varphi_{j+1}=\eta_{j+1}$,
and 
$\displaystyle \{\varphi_j, \varphi_k\}=\varphi_j\varphi_k+\varphi_k\varphi_j=\delta_{jk}$ 
for all $j$, $k$.
Hence, the Hamiltonian is written 
as a sum of two-body products of fermion operators $\varphi_{j}$. 

The transformation (\ref{transmain}) is generated from $\{\eta_j\}$, 
and only the commutation relation (\ref{cond2}) is needed 
to obtain the free energy. 
When we consider the transverse Ising chain, i.e. 
$\eta_{2j-1}=\sigma_{j}^z$ and $\eta_{2j}=\sigma_{j}^x\sigma_{j+1}^x$, 
(\ref{transmain}) reduces to the original Jordan-Wigner transformation. 
In other cases, we obtain other transformations 
that diagonalize the Hamiltonian. 

This fermionization formula was applied to 
one-dimensional quantum spin chains \cite{17Minami}\cite{20Yanagihara},
and the honeycomb-lattice Kitaev model 
with the Wen-Toric code interactions \cite{19Minami}. 
The key idea of this transformation was developed 
into a graph-theoretic treatment \cite{20Ogura}, 
in which the transformations of operators 
are expressed as modifications of graphs, 
and the kernel of its adjacency matrices 
corresponds to conserved quantities of the system.
The condition (\ref{cond2}) was independently considered 
to introduce models that can be mapped to the free fermion system, 
and investigated in terms of the graph theory 
\cite{20ChapmanFlammia}
\cite{20ElmanChapmanFlammia}.
\\

In this short note, 
we extend the Onsager's result 
and show that there exist an infinite number of interactions 
that satisfy the Dolan-Grady condition, 
and hence exist an infinite number of realizations 
of (quotients of)  the Onsager algebra. 
We also consider the operators that satisfy the conditions 
\begin{eqnarray}
&&
\eta_{j}\eta_{j+1}=\omega\eta_{j+1}\eta_{j},
\hspace{0.6cm}
\eta_{j}\eta_{k}=\eta_{k}\eta_{j}
\hspace{0.3cm}
(|j-k|>1),
\nonumber
\\
&&
\eta_{j}^{n}=1, 
\hspace{0.6cm}
\omega=e^{i\frac{2\pi}{n}},
\label{condn}
\end{eqnarray}
and show 
an infinite number of interactions 
satisfying the Dolan-Grady condition. 
Note that the replacement 
$\omega\mapsto\omega^{-1}$ 
corresponds to the inversion of the indices 
$j\mapsto M-j+1$. 

Throughout this short note, 
the cyclic boundary condition 
\begin{eqnarray}
\eta_{M+j}=\eta_{j},
\label{cyclic}
\end{eqnarray}
where $M$ is the number of operators, is assumed. 
The conditions (\ref{cond2}) or (\ref{condn}) 
together with the cyclic boundary condition (\ref{cyclic}) 
yield the Theorem 1-3 and Corollary 1.

When $n=2$, the condition (\ref{condn}) reduces to (\ref{cond2}). 
Operators that satisfy (\ref{condn}) 
were considered in \cite{80FradkinKadanoff}, 
and also investigated in \cite{83HowesKadanoffDebNijs} with $n=3$ 
concerning the incommensurate phase, 
and considered with arbitrary integer $n$ in \cite{85GehlenRittenberg}. 
The Hamiltonian in \cite{85GehlenRittenberg} 
was later obtained from the transfer matrix 
of the two-dimensional chiral Potts model 
\cite{87AuYangMcCoyPerkTangYan}
\cite{88BaxterPerkAuYang}. 
The Baxter's clock chain 
\cite{89Baxter_PhysLettA}
\cite{89Baxter_JStatPhys61} 
can also be written in terms of the operators that satisfy (\ref{condn}),  
and can be rewritten, 
through the Fradkin-Kadanoff transformation \cite{80FradkinKadanoff}, 
in terms of the parafermions, 
which are $Z_{n}$ generalizations of Majorana fermions. 
It is easily shown that 
the Fradkin-Kadanoff transformation can be obtained from the formula (\ref{transmain}).  
It is now known that 
the Baxter's clock chain can be regarded as 'free' parafermions 
\cite{12Fendley}
\cite{14Fendley}. 
Generalizations of the relation (\ref{condn}) 
were investigated in \cite{14Fendley}, 
and in \cite{20AlcarazPimenta}-\cite{20AlcarazPimenta2}, 
and the corresponding Hamiltonians 
were shown to have an Ising-like spectrum and be integrable.

In Theorem 1-3 and Corollary 1, 
we will show that 
operators composed of $\eta_{j}$'s 
satisfy the Dolan-Grady condition.  
Only the algebraic relations are needed to derive the results, 
and hence any operators that satisfy (\ref{cond2}) or (\ref{condn}) 
generate operators 
that satisfy the Dolan-Grady condition, 
and hence generate a quotient of the Onsager algebra. 
In Table 1 and Theorem 5, 
we show explicit examples of operators that satisfy (\ref{condn}), 
including the interactions of the transverse Ising chain, 
and so-called the superintegrable chiral Potts model. 
At last, we will comment on a fact that 
an infinite number of models with inhomogeneous interactions 
also become integrable. 
\\


\section{Onsager algebra and Theorems}
Let us consider series of operators $\{A_{j}\}$ and $\{G_{j}\}$, 
where $j\in {\bf Z}$. 
The Onsager algebra is a Lie algebra 
defined via the relations 
\begin{eqnarray}
[A_j,A_k]=4G_{j-k},
\hspace{0.4cm}
[G_m,A_l]=2A_{l+m}-2A_{l-m},
\hspace{0.4cm}
[G_j,G_k]=0.
\label{Onsagerrel}
\end{eqnarray}
It is known that a pair of operators, $A_{0}$ and $A_{1}$, 
recursively generate all the $A_{j}$ and $G_{k}$ in (\ref{Onsagerrel}) 
provided that they satisfy the relations 
\begin{eqnarray}
[\:A_{0}[\:A_{0}[\:A_{0}, A_{1}\:]\:]\:]
&=&
C[\:A_{0}, A_{1}\:],
\label{DG1}
\\
\:
[\:A_{1}[\:A_{1}[\:A_{1}, A_{0}\:]\:]\:]
&=&
C[\:A_{1}, A_{0}\:],
\label{DG2}
\end{eqnarray}
where $C$ is a constant. 
We call (\ref{DG1}) and (\ref{DG2}) the Dolan-Grady condition. 

When we consider a Hamiltonian 
\begin{eqnarray}
{\cal H}
=
A_{0}+kA_{1}, 
\label{HamA0kA1}
\end{eqnarray}
where $k$ is a constant, 
it can be derived that ${\cal H}$ 
belongs to an infinite family of mutually commuting operators, 
i.e. ${\cal H}$ is integrable.  
For irreducible finite dimensional representations of the Onsager algebra, 
the general form of the eigenvalue 
was obtained in \cite{90Davies} as 
\begin{eqnarray*}
(\alpha+\beta k)+\sum_{j=1}^{n}4m_{j}\sqrt{1+k^{2}+2k\cos\theta_{j}},
\hspace{0.4cm}
m_{j}=0, \pm 1, \pm 2, \ldots, \pm s_{j},
\end{eqnarray*}
where $\alpha$ and $\beta$ are constants, 
and $s_{j}$ are positive integers.

Onsager introduced, 
for the purpose to solve the rectangular Ising model, 
the transfer matrix which is expressed by the operators 
\begin{eqnarray}
A_{0}=\sum_{j=1}^{N}\sigma_{j}^{x}, 
\hspace{0.4cm}
A_{1}=\sum_{j=1}^{N}\sigma_{j}^{z}\sigma_{j+1}^{z}. 
\label{A0A1Ising}
\end{eqnarray}
These $A_{0}$ and $A_{1}$ satisfy the Dolan-Grady condition, 
and in this case (\ref{HamA0kA1}) 
is the Hamiltonian of the transverse Ising chain \cite{17Perk}. 
\\


\begin{theorem}
Let us introduce 
\begin{eqnarray}
A_{0}
=
\sum_{j=1}^{N}\eta_{2j-1},
\hspace{0.6cm}
A_{1}
=
\sum_{j=1}^{N}\eta_{2j}, 
\end{eqnarray}
where 
$N\geq 2$, and $\eta_{j}$ satisfy (\ref{cond2}) and (\ref{cyclic}). 
Then $A_{0}$ and $A_{1}$ 
satisfy the Dolan-Grady condition 
(\ref{DG1}) and (\ref{DG2}) with $C=16$. 
\end{theorem}

Direct calculations yield Theorem 1. 
We can also convince 
\begin{eqnarray*}
&&
A_{2}
=
\sum_{j=1}^{N}\eta_{2j}\eta_{2j+1}\eta_{2j+2},
\hspace{1.2cm}
A_{3}
=
\sum_{j=1}^{N}\eta_{2j}\eta_{2j+1}\eta_{2j+2}\eta_{2j+3}\eta_{2j+4},
\\
&&
A_{-1}
=
\sum_{j=1}^{N}\eta_{2j-3}\eta_{2j-2}\eta_{2j-1},
\hspace{0.6cm}
A_{-2}
=
\sum_{j=1}^{N}\eta_{2j-5}\eta_{2j-4}\eta_{2j-3}\eta_{2j-2}\eta_{2j-1},
\\
&&
G_{0}
=
0,
\hspace{1.2cm}
G_{1}
=
\frac{1}{2}\sum_{j=1}^{N}
(\eta_{2j}\eta_{2j+1}-\eta_{2j-1}\eta_{2j}),
\\
&&
G_{2}
=
\frac{1}{2}\sum_{j=1}^{N}
(\eta_{2j}\eta_{2j+1}\eta_{2j+2}\eta_{2j+3}-\eta_{2j-1}\eta_{2j}\eta_{2j+1}\eta_{2j+2}),
\\
&&
G_{3}
=
\frac{1}{2}\sum_{j=1}^{N}
(\eta_{2j}\eta_{2j+1}\eta_{2j+2}\eta_{2j+3}\eta_{2j+4}\eta_{2j+5}-\eta_{2j-1}\eta_{2j}\eta_{2j+1}\eta_{2j+2}\eta_{2j+3}\eta_{2j+4}),
\end{eqnarray*}
and generally
\begin{eqnarray*}
&&
A_{l}
=
\sum_{j=1}^{N}\eta_{2j}\eta_{2j+1}\cdots\eta_{2j+2l-2},
\hspace{0.9cm}
A_{-l}
=
\sum_{j=1}^{N}\eta_{2j-2l-1}\eta_{2j-2l}\cdots\eta_{2j-1},
\end{eqnarray*}
and
\begin{eqnarray*}
&&
G_{l}
=
\frac{1}{2}\sum_{j=1}^{N}
(\eta_{2j}\eta_{2j+1}\eta_{2j+2}\cdots\eta_{2j+2l-1}-\eta_{2j-1}\eta_{2j}\eta_{2j+1}\cdots\eta_{2j+2l-2}).
\end{eqnarray*}

It is straightforward to derive that 
$A_{l}$ and $G_{l}$ satisfy $A_{l+2N}=A_{l}$ and $G_{l+2N}=G_{l}$.  
The relation $A_{l+2N}=A_{l}$ 
is the closure relation considered by Davies in [8]. 
This means that the algebra generated from $A_{0}$ and $A_{1}$ in Theorem 1 
is a quotient of the Onsager algebra, 
i.e. an Onsager algebra with the restriction $A_{l+2N}=A_{l}$.

Theorem 1 shows existence of an infinite number of Hamiltonians 
that are expressed by (\ref{HamA0kA1}) 
and hence governed by the Onsager algebra, 
because we know examples of operators that satisfy (\ref{cond2}), 
such as 
the interactions of the transverse Ising chain: 
$\eta^{(1)}_{2j-1}=\sigma_{j}^z$ and 
$\eta^{(1)}_{2j}=\sigma_{j}^x\sigma_{j+1}^x$, 
those of the Kitaev chain: 
$\eta^{(2)}_{2j-1}=\sigma_{2j-1}^x\sigma_{2j}^x$ and 
$\eta^{(2)}_{2j}=\sigma_{2j}^y\sigma_{2j+1}^y$, 
those of the cluster model: 
$\eta^{(3)}_{2j-1}=\sigma_{2j-1}^x\sigma_{2j}^z\sigma_{2j+1}^x$ and 
$\eta^{(3)}_{2j}=\sigma_{2j}^x1_{2j+1}\sigma_{2j+2}^x$, 
and other infinite number of interactions 
listed in Table 1 and Table 2 given in \cite{17Minami}.


We will consider Theorem 1 in the cases 
where the condition (\ref{condn}) is satisfied. 
Examples of operators that satisfy (\ref{condn}) 
are shown in Table 1,  
where operators $\eta_{j}$ 
form one or several series of operators;  
operators in each series satisfy (\ref{condn}), 
and operators from different series commute with each other. 
The operators $Z$, $X$ and $Y$ in Table 1 are defined, 
with $\omega^{n}=1$, as 
\begin{eqnarray}
&&
Z
=
\left(
\begin{array}{ccccc}
1&&&&\\
&\omega&&&\\
&&\omega^{2}&&\\
&&&\ddots&\\
&&&&\omega^{n-1}
\end{array}
\right), 
\hspace{0.4cm}
X
=
\left(
\begin{array}{ccccc}
0&&&&1\\
1&0&&&\\
&1&0&&\\
&&&\ddots&\\
&&&1&0
\end{array}
\right), 
\nonumber\\
&&
Y
=
\omega^{-\frac{1}{2}(n-1)}
\left(
\begin{array}{ccccc}
0&\omega^{n-2}&&&\\
&0&\omega^{n-3}&&\\
&&0&\ddots&\\
&&&\ddots&1\\
\omega^{n-1}&&&&0
\end{array}
\right),
\end{eqnarray}
and $Z_{j}$, $X_{j}$ and $Y_{j}$ are defined as 
\begin{eqnarray}
Q_{j}
=
1\otimes\cdots\otimes 1\otimes\stackrel{\stackrel{j}{\vee}}{Q}\otimes 1\otimes\cdots\otimes 1,
\hspace{0.6cm}
Q=Z, X, Y.
\end{eqnarray}
The operators $Z$, $X$, and $Y$ satisfy 
$ZX=\omega XZ$, 
$XY=\omega YX$, and 
$YZ=\omega ZY$. 
Then we can derive the following corollary.


\begin{corollary}
Let us consider 
\begin{eqnarray}
A_{0}
=
\sum_{j=1}^{N}\eta_{2j-1}^{k},
\hspace{0.6cm}
A_{1}
=
\sum_{j=1}^{N}\eta_{2j}^{l}, 
\end{eqnarray}
where $N\geq 2$, 
and $n$ is even, and $k=l=n/2$, 
and $\eta_{j}$ satisfy (\ref{condn}) and (\ref{cyclic}). 
Then $A_{0}$ and $A_{1}$ satisfy the Dolan-Grady condition 
(\ref{DG1}) and (\ref{DG2}) with $C=16$ 
when $n/2$ is odd, 
and $[A_{0}, A_{1}]=0$ 
when $n/2$ is even. 
\end{corollary}

\noindent
{\rm Proof:} 
Let $\zeta_{j}=\eta_{j}^{k}$, then $\zeta_{j}^{2}=1$. 
We find  
$\zeta_{j}\zeta_{j+1}
=\eta_{j}^{k}\eta_{j+1}^{k}
=\omega^{k^{2}}\eta_{j+1}^{k}\eta_{j}^{k}
=\omega^{k^{2}}\zeta_{j+1}\zeta_{j}$, 
where 
$\omega^{k^{2}}=e^{i\frac{2\pi}{n}k^{2}}=e^{i\pi k}=(-1)^{k}$. 
If $k$ is even, then $[A_{0}, A_{1}]=0$.  
If $k$ is odd, 
the operators $A_{0}$ and $A_{1}$ written in terms of $\zeta_{j}$ 
satisfy the Dolan-Grady condition (\ref{DG1}) and (\ref{DG2}) 
since $\zeta_{j}\zeta_{j+1}=-\zeta_{j+1}\zeta_{j}$ 
and $\zeta_{j}^{2}=1$ 
\qed
\\

When $n=2$, 
then $A_{0}=\sum_{j=1}^{N}\eta_{2j-1}$ and $A_{1}=\sum_{j=1}^{N}\eta_{2j}$, 
and thus Theorem 1 is obtained from Corollary 1.

In Corollary 1, the relation 
$\eta_{j}\eta_{j+1}=\omega\eta_{j+1}\eta_{j}$
is assumed. 
When we assume 
$\eta_{j}\eta_{j+1}=\omega_{j}\eta_{j+1}\eta_{j}$, 
where $\omega_{j}$ depends on $j$ and equals to $\omega$ or $\omega^{-1}$, 
we obtain 
$\zeta_{j}\zeta_{j+1}
=\omega^{\pm k^{2}}\zeta_{j+1}\zeta_{j}
=-\zeta_{j+1}\zeta_{j}$, 
and the Dolan-Grady condition (\ref{DG1}) and (\ref{DG2}) 
is satisfied again. 
\\


\begin{theorem}
Let us consider 
\begin{eqnarray}
A_{0}
=
\sum_{j=1}^{N}(\eta_{2j-1}^{k}-\eta_{2j-1}^{n-k}),
\hspace{0.6cm}
A_{1}
=
\sum_{j=1}^{N}(\eta_{2j}^{k}-\eta_{2j}^{n-k}),
\end{eqnarray}
where $N\geq 2$, 
and $k=n/3$ is an integer that satisfy $1\leq k\leq n-1$, 
and $\eta_{j}$ satisfy (\ref{condn}) and (\ref{cyclic}). 
Then $A_{0}$ and $A_{1}$ satisfy the Dolan-Grady condition 
(\ref{DG1}) and (\ref{DG2}) with $C=-27$ 
when $k=3m-1, 3m-2$ $\:(m\in{\bf N})$, 
and $[A_{0}, A_{1}]=0$ 
when $k=3m$. 
\end{theorem}

It is easy to show 
\begin{eqnarray}
\:[\eta_{2j-1}^{k}, \eta_{2j}^{l}]
&=&
\eta_{2j-1}^{k}\eta_{2j}^{l}-\eta_{2j}^{l}\eta_{2j-1}^{k}
=
(1-\omega^{-kl})\eta_{2j-1}^{k}\eta_{2j}^{l}, 
\nonumber
\\ 
\:[\eta_{2j+1}^{k}, \eta_{2j}^{l}]
&=&
\eta_{2j+1}^{k}\eta_{2j}^{l}-\eta_{2j}^{l}\eta_{2j+1}^{k}
=
(1-\omega^{kl})\eta_{2j+1}^{k}\eta_{2j}^{l}.
\end{eqnarray}
The inner derivatives in terms of $\eta_{2j\mp 1}^{k}$, 
operated to $\eta_{2j}^{l}$, 
are therefore equivalent to the multiplications of 
$(1-\omega^{\mp kl})\eta_{2j\mp 1}^{k}$ 
from the left. 
Then we will prove the Theorem. 
\\

\noindent
{\rm Proof:} 
It is easy to show 
\begin{eqnarray}
[\sum_{i=1}^{N}\eta_{2i-1}^{k}, \eta_{2j}^{k}]
&=&
[\eta_{2j-1}^{k}, \eta_{2j}^{k}]+[\eta_{2j+1}^{k}, \eta_{2j}^{k}]
\nonumber
\\
&=&
\Big((1-\omega^{-k^{2}})\eta_{2j-1}^{k}+(1-\omega^{k^{2}})\eta_{2j+1}^{k}\Big)
\eta_{2j}^{k}
\nonumber
\\
&=&
a_{11j}\eta_{2j}^{k},
\\
\:[\sum_{i=1}^{N}\eta_{2i-1}^{n-k}, \eta_{2j}^{k}]
&=&
[\eta_{2j-1}^{n-k}, \eta_{2j}^{k}]+[\eta_{2j+1}^{n-k}, \eta_{2j}^{k}]
\nonumber
\\
&=&
\Big((1-\omega^{k^{2}})\eta_{2j-1}^{n-k}+(1-\omega^{-k^{2}})\eta_{2j+1}^{n-k}\Big)
\eta_{2j}^{k}
\nonumber
\\
&=&
a_{21j}\eta_{2j}^{k},
\end{eqnarray}
where 
\begin{eqnarray}
a_{11j}&=&z\eta_{2j-1}^{k}+{\bar z}\eta_{2j+1}^{k},
\hspace{0.6cm}
a_{21j}={\bar z}\eta_{2j-1}^{n-k}+z\eta_{2j+1}^{n-k},
\nonumber
\\
z&=&1-\omega^{-k^{2}},
\hspace{2.2cm}
{\bar z}=1-\omega^{k^{2}}.
\end{eqnarray}
Then we obtain 
\begin{eqnarray}
[A_{0}, \eta_{2j}^{k}]
&=&
\Big((z\eta_{2j-1}^{k}+{\bar z}\eta_{2j+1}^{k})
-({\bar z}\eta_{2j-1}^{n-k}+z\eta_{2j+1}^{n-k})\Big)\eta_{2j}^{k}
\label{A1etak}
\\
&=&
(a_{11j}-a_{21j})\eta_{2j}^{k},
\nonumber
\end{eqnarray}
Similarly, we obtain 
\begin{eqnarray}
[A_{0}, \eta_{2j}^{n-k}]
&=&
\Big(({\bar z}\eta_{2j-1}^{k}+z\eta_{2j+1}^{k})
-(z\eta_{2j-1}^{n-k}+{\bar z}\eta_{2j+1}^{n-k})\Big)\eta_{2j}^{n-k}
\label{A1etank}
\\
&=&
(a_{12j}-a_{22j})\eta_{2j}^{n-k},
\nonumber
\end{eqnarray}
where 
\begin{eqnarray}
a_{12j}={\bar z}\eta_{2j-1}^{k}+z\eta_{2j+1}^{k},
\hspace{0.6cm}
a_{22j}=z\eta_{2j-1}^{n-k}+{\bar z}\eta_{2j+1}^{n-k}.
\end{eqnarray}
Since $\eta_{j}$'s with odd $j$ commute with each other, 
we obtain 
\begin{eqnarray}
[A_{0}, [A_{0}, [A_{0}, \eta_{2j}^{k}]]]
=
(a_{11j}-a_{21j})^{3}\eta_{2j}^{k},
\label{A1A1A1etak}
\end{eqnarray}
and 
\begin{eqnarray}
[A_{0}, [A_{0}, [A_{0}, \eta_{2j}^{n-k}]]]
=
(a_{12j}-a_{22j})^{3}\eta_{2j}^{n-k}.
\end{eqnarray}
We find the following terms appear in (\ref{A1A1A1etak}) 
\begin{eqnarray}
a_{11j}^{3}
&=&
z^{3}\eta_{2j-1}^{3k}+3z^{2}{\bar z}\eta_{2j-1}^{2k}\eta_{2j+1}^{k}
+3z{\bar z}^{2}\eta_{2j-1}^{k}\eta_{2j+1}^{2k}+{\bar z}^{3}\eta_{2j+1}^{3k},
\nonumber
\\
a_{21j}^{3}
&=&
{\bar z}^{3}\eta_{2j-1}^{-3k}+3{\bar z}^{2}z\eta_{2j-1}^{-2k}\eta_{2j+1}^{-k}
+3{\bar z}z^{2}\eta_{2j-1}^{-k}\eta_{2j+1}^{-2k}+z^{3}\eta_{2j+1}^{-3k},
\end{eqnarray}
and 
\begin{eqnarray}
a_{11j}^{2}a_{21j}
&=&
3z^{2}{\bar z}\eta_{2j-1}^{k}+3z{\bar z}^{2}\eta_{2j+1}^{k}
+z^{3}\eta_{2j-1}^{2k}\eta_{2j+1}^{-k}
+{\bar z}^{3}\eta_{2j-1}^{-k}\eta_{2j+1}^{2k},
\label{a112a21}
\\
a_{11j}a_{21j}^{2}
&=&
3z{\bar z}^{2}\eta_{2j-1}^{-k}+3z^{2}{\bar z}\eta_{2j+1}^{-k}
+z^{3}\eta_{2j-1}^{k}\eta_{2j+1}^{-2k}
+{\bar z}^{3}\eta_{2j-1}^{-2k}\eta_{2j+1}^{k}.
\label{a11a212}
\end{eqnarray}
The assumption $n=3k$ yields $\eta_{2j-1}^{3k}=1$, $\eta_{2j+1}^{3k}=1$, 
and hence $a_{11j}^{3}-a_{21j}^{3}=0$. 
From $3k=n$, we find $\omega^{3k^{2}}=(\omega^{n})^{k}=1$ and ${\bar z}^{3}+z^{3}=0$, 
then the last two terms in the right-hand side of (\ref{a112a21}), 
and also those of (\ref{a11a212}), 
cancel each other, respectively.   
Together with (\ref{A1etak}), we obtain 
\begin{eqnarray}
[A_{0}, [A_{0}, [A_{0}, \eta_{2j}^{k}]]]
=
-9z{\bar z}[A_{0}, \eta_{2j}^{k}].
\label{A1A1A1A0A1etak}
\end{eqnarray}
Similarly we find 
\begin{eqnarray}
[A_{0}, [A_{0}, [A_{0}, \eta_{2j}^{n-k}]]]
=
-9z{\bar z}[A_{0}, \eta_{2j}^{n-k}].
\label{A1A1A1A0A1etank}
\end{eqnarray}
When $k=3m$, then $\omega^{k^{2}}=\omega^{3m\cdot k}=(\omega^{n})^{m}=1$, $z=0$, 
and from (\ref{A1etak}) and (\ref{A1etank}) we find $[A_{0}, A_{1}]=0$. 
When $k=3m-1$, then 
$\omega^{k^{2}}=\omega^{3k\cdot m}\omega^{-k}
=(\omega^{n})^{m}(\omega^{k})^{-1}
=1\cdot(e^{i\frac{2\pi}{n}\frac{n}{3}})^{-1}=e^{-i\frac{2\pi}{3}}$, 
and when $k=3m-2$, then 
$\omega^{k^{2}}=\omega^{3k\cdot m}\omega^{-2k}
=(\omega^{n})^{m}(\omega^{k})^{-2}
=1\cdot(e^{i\frac{2\pi}{n}\frac{n}{3}})^{-2}=e^{i\frac{2\pi}{3}}$. 
We thus obtain $z{\bar z}=3$, $[A_{0}, A_{1}]\neq 0$, 
and the first part of the Dolan-Grady condition (\ref{DG1}) is satisfied 
with $C=-9z{\bar z}$.
The second part of the Dolan-Grady condition (\ref{DG2}) 
is obtained by the shift of indices $2j\mapsto 2j+1$.
\qed
\\

Let $\zeta_{j}=\eta_{j}^{k}$ $(k=n/3)$. 
Then we find  
$\zeta_{j}\zeta_{j+1}
=\eta_{j}^{k}\eta_{j+1}^{k}
=\omega^{k^{2}}\eta_{j+1}^{k}\eta_{j}^{k}
=\omega^{k^{2}}\zeta_{j+1}\zeta_{j}$. 
When $k=3m$, then 
$\omega^{k^{2}}=1$, 
and therefore 
$\zeta_{j}\zeta_{j+1}=\zeta_{j+1}\zeta_{j}$, 
and we obtain $[A_{0}, A_{1}]=0$. 
When $k=3m-2$, then 
$\omega^{k^{2}}=e^{i\frac{2\pi}{3}}$, 
and we find $\zeta_{j}\zeta_{j+1}=e^{i\frac{2\pi}{3}}\zeta_{j+1}\zeta_{j}$.  
When $k=3m-1$, then 
$\omega^{k^{2}}=e^{-i\frac{2\pi}{3}}$, 
and 
${\bar \zeta}_{j}{\bar \zeta}_{j+1}
=e^{i\frac{2\pi}{3}}{\bar \zeta}_{j+1}{\bar \zeta}_{j}$, 
where ${\bar \zeta}_{j}=\zeta_{2N-j+1}$.   
The case with $\omega=e^{i\frac{2\pi}{3}}$ 
was already considered in \cite{12FjelstadMansson} 
though the derivation is different. 

With the choice of the operators 
$\displaystyle \zeta_{2j-1}=X_{j}$ and 
$\displaystyle \zeta_{2j}=Z_{j}Z_{j+1}^{\dag}$, 
the Hamiltonian $A_{0}+kA_{1}$ with $n=3$ can be written as 
${\cal H}_{B}-{\cal H}_{B}^{\dag}$, 
where ${\cal H}_{B}$ is the Baxter's clock model 
\cite{89Baxter_PhysLettA}\cite{89Baxter_JStatPhys61}. 
\\


\begin{theorem}
Let us consider 
\begin{eqnarray}
A_{0}=
\sum_{j=1}^{N}\sum_{k=1}^{n-1}
\frac{\eta_{2j-1}^{k}}{1-\omega^{-k}}, 
\hspace{0.6cm}
A_{1}=
\sum_{j=1}^{N}\sum_{k=1}^{n-1}
\frac{\eta_{2j}^{k}}{1-\omega^{-k}},
\end{eqnarray}
where $N\geq 2$, 
and $\eta_{j}$ together with $\omega$ satisfy (\ref{condn}), 
and the cyclic boundary condition (\ref{cyclic}) is assumed. 
Then $[A_{0}, A_{1}]\neq 0$, 
and $A_{0}$, $A_{1}$ satisfy the Dolan-Grady condition 
(\ref{DG1}) and (\ref{DG2}) with $C=n^{2}$. 
\end{theorem}

For the purpose to prove this Theorem, 
we use the formula \cite{75Hansen}\cite{85GehlenRittenberg}
\begin{eqnarray}
\sum_{l=1}^{n-1}\frac{\omega^{\frac{1}{2}(m-1)l}}{1-\omega^{-l}}
=
\frac{1}{2}(n-m),
\hspace{0.6cm}
\sum_{l=1}^{n-1}\frac{\omega^{-\frac{1}{2}(m+1)l}}{1-\omega^{-l}}
=
-\frac{1}{2}(n-m),
\label{sumformula}
\\
n=2, 3, 4, 5, \ldots,\hspace{0.4cm}
m=1, 3, 5, \ldots
\nonumber
\\
m\leq n \hspace{0.2cm}{\rm and}\hspace{0.2cm}\omega=e^{i\frac{2\pi}{n}}.
\nonumber
\end{eqnarray}
The first formula is equivalent to (2.16) of \cite{85GehlenRittenberg}, 
and the second is obtained from the first. 
\\

\noindent
{\bf Proof:}
\begin{eqnarray}
[A_{0}, \eta_{2j}^{k}]
&=&
[\sum_{i=1}^{N}\sum_{l=1}^{n-1}\frac{\eta_{2i-1}^{l}}{1-\omega^{-l}}, \eta_{2j}^{k}]
\nonumber
\\
&=&
\sum_{l=1}^{n-1}\frac{1}{1-\omega^{-l}}[\sum_{i=1}^{N}\eta_{2i-1}^{l}, \eta_{2j}^{k}]
\nonumber
\\
&=&
\sum_{l=1}^{n-1}\frac{1}{1-\omega^{-l}}
\Big(
(1-\omega^{-kl})\eta_{2j-1}^{l}+(1-\omega^{kl})\eta_{2j+1}^{l}
\Big)
\eta_{2j}^{k}
\nonumber
\\
&=&
\Big(
\sum_{l=1}^{n-1}c_{l}(kl)\eta_{2j-1}^{l}+\sum_{l=1}^{n-1}c_{l}(-kl)\eta_{2j+1}^{l}
\Big)
\eta_{2j}^{k},
\label{DolanGrady1-3}
\end{eqnarray}
where 
\begin{eqnarray}
c_{l}(m)=\frac{1-\omega^{-m}}{1-\omega^{-l}}.
\end{eqnarray}
The Dolan-Grady relation (\ref{DG1}) is satisfied if 
\begin{eqnarray}
\Big(\Delta_{k}(\eta_{2j-1}, \eta_{2j+1})^{3}-C\Delta_{k}(\eta_{2j-1}, \eta_{2j+1})\Big)
\eta_{2j}^{k}
=0,
\label{DGThe3-1}
\end{eqnarray}
where
\begin{eqnarray}
\Delta_{k}(x, y)
=
\Delta_{k}(x)+\Delta_{-k}(y),
\hspace{0.6cm}
\Delta_{k}(x)
=
\sum_{l=1}^{n-1}c_{l}(kl)x^{l}.
\end{eqnarray}
For the purpose to derive (\ref{DGThe3-1}), 
it is sufficient to show that  
\begin{eqnarray}
\Delta_{k}(x, y)^{3}-n^{2}\Delta_{k}(x, y)=0 
\label{DGThe3-2}
\end{eqnarray}
as a polynomial, 
with the condition $x^n=1$ and $y^n=1$.  
For the purpose to show (\ref{DGThe3-2}), 
it is sufficient to show that (\ref{DGThe3-2}) is satisfied 
with independent numbers 
$x, y=1$, $\omega, \omega^2, \ldots, \omega^{n-1}$, 
where $\omega=e^{i\frac{2\pi}{n}}$ (that satisfy $\omega^n=1$). 
It is straightforward to show, 
with the use of (\ref{sumformula}), 
which is valid when $m\leq n$, 
that 
$\Delta_{k}(\omega^{s})=-k \:\:(k=1, 2,\ldots, s)$, 
$\Delta_{k}(\omega^{s})=-(n-k)\:\:(k=s+1, \ldots, n-1)$, 
and 
$\Delta_{-k}(\omega^{s})=-(n-k) \:\:(k=1, 2,\ldots, s)$, 
$\Delta_{-k}(\omega^{s})=k\:\:(k=s+1, \ldots, n-1)$, 
and thus $\Delta_{k}(x, y)$ takes $n$ or $0$ or $-n$, 
which yields (\ref{DGThe3-1}) with $C=n^{2}$ 
and $[A_{0}, A_{1}]\neq 0$. 
Similarly we obtain (\ref{DG2})
\qed 
\\

With the choice of the operators 
$\displaystyle \eta_{2j-1}=Z_{j}$ and 
$\displaystyle \eta_{2j}=X_{j}X_{j+1}^{\dag}$, 
the Hamiltonian $A_{0}+kA_{1}$ results in 
that of the superintegrable chiral Potts chain \cite{85GehlenRittenberg}.
Note that in \cite{85GehlenRittenberg}, 
the Dolan-Grady condition was derived 
with the use of the explicit matrix representation. 
In our derivation, Theorem 4 is proved 
using only the algebraic relations, 
and thus valid for all operators 
which satisfy (\ref{condn}).

The operators $A_{0}$ and $A_{1}$ in Theorem 1 
satisfy a secular equation 
and the generated algebra is a quotient of the Onsager algebra. 
It is an important problem to find the secular equations 
in the cases of Theorem 2 and 3, 
and specify the possible form of the eigenvalue for all cases. 
\\


We can find in Table 1 
a list of operators that satisfy (3). 
The next Theorem shows that 
once we find an set of operators satisfying (3), 
we can generate other sets of operators that satisfy (3).
\begin{theorem}
Let us consider the case 
$\displaystyle \eta_{j}=\prod_{k=1}^{N}(X_{k}^{x_{jk}}Z_{k}^{z_{jk}})$, 
where $x_{jk}$ and $z_{jk}$ are non-negative integers. 
Let $\varphi_{X}$ and $\varphi_{Z}$ be transformations 
defined by 
\begin{eqnarray}
\varphi_{Z}
&:&
X_{k}\mapsto X_{k}, 
\hspace{0.8cm}
Z_{k}\mapsto Z_{k}^{-1}, 
\nonumber
\\
\varphi_{X}
&:&
X_{k}\mapsto X_{k}^{-1}, 
\hspace{0.6cm}
Z_{k}\mapsto Z_{k}
\label{transXZ}
\end{eqnarray}
for all $k$, and $\varphi_{XZ}=\varphi_{X}\circ\varphi_{Z}$. 
Assume that the operators $\{\eta_{j}\}=\{\eta_{2j-1}\}\cup\{\eta_{2j}\}$ 
satisfy the condition (\ref{condn}), 
then the set of operators
\begin{eqnarray}
&&
\{\varphi_{Z}(\eta_{2j-1})\}\cup\{\varphi_{Z}(\eta_{2j})\}, 
\hspace{0.6cm}
\{\varphi_{X}(\eta_{2j-1})\}\cup\{\varphi_{X}(\eta_{2j})\}, 
\nonumber
\\
&&
\{\varphi_{XZ}(\eta_{2j-1})\}\cup\{\eta_{2j}\}, 
\hspace{0.4cm}
\{\eta_{2j-1}\}\cup\{\varphi_{XZ}(\eta_{2j})\}, 
\label{transXZ5}
\\
&&
\{\varphi_{XZ}(\eta_{2j-1})\}\cup\{\varphi_{XZ}(\eta_{2j})\}
\nonumber
\end{eqnarray}
also satisfy the condition (\ref{condn}). 
\end{theorem}

\noindent
{\bf Proof:}
The first condition in (\ref{condn}), 
$\eta_{j}\eta_{j+1}=\omega\eta_{j+1}\eta_{j}$ or 
$\eta_{j}\eta_{j+1}=\omega^{-1}\eta_{j+1}\eta_{j}$, 
 is written as 
\begin{eqnarray}
\sum_{k=1}^{N}(z_{j k}x_{j+1 k}-x_{j k}z_{j+1 k})
&=&
1
\hspace{0.4cm}
{\rm or}
\hspace{0.4cm}
-1.
\label{condnel1}
\end{eqnarray}
The second condition in (\ref{condn}), 
$\eta_{i}\eta_{j}=\eta_{j}\eta_{i}$ $(|i-j|>1)$, 
is written as
\begin{eqnarray}
\sum_{k=1}^{N}(z_{i k}x_{j k}-x_{i k}z_{j k})
&=&
0
\hspace{0.6cm}
|i-j|>1.
\label{condnel2}
\end{eqnarray}
The transformations (\ref{transXZ}) result in 
\begin{eqnarray}
\varphi_{Z}:
z_{jk}\mapsto -z_{jk}, 
\hspace{0.6cm}
\varphi_{X}:
x_{jk}\mapsto -x_{jk}  
\end{eqnarray}
for all $k$. 
Then it is easy to convince 
that the conditions (\ref{condnel1}) and (\ref{condnel2}) 
are also satisfied after the transformations (\ref{transXZ5})
\qed
\\


Uglov anf Ivanov \cite{95UglovIvanov} 
considered a generalization of the original Onsager algebra. 
They considered a Hamiltonian of the form 
\begin{eqnarray}
-\beta{\cal H}
=
\sum_{j=1}^{N}K_{j}e_{j}
\hspace{0.6cm}
(N\geq 3)
\label{HamRdm}
\end{eqnarray}
and derived that if the operators $e_{j}$ satisfy the relation 
\begin{eqnarray}
[e_{i}, [e_{i}, e_{j}]]&=&e_{j}
\hspace{0.6cm}
(|i-j|=1),
\nonumber\\ \:
[e_{i}, e_{j}]&=&0
\hspace{0.7cm}
(|i-j|>1),
\label{UIcond}
\end{eqnarray}
then there exists an infinite family of integrals 
$\{I_{m}\}$, 
where $I_{0}={\cal H}$ 
and $[I_{m}, I_{n}]=0$ $(m, n\geq 1)$. 

We would like to note that 
$\displaystyle \frac{1}{2}\eta_{j}^{k}$ 
($k=n/2=$odd) 
satisfy the condition (\ref{UIcond}). 
A list of operators $\eta_{j}$ with $n=2$ 
can be found in Table.1 of \cite{17Minami}, 
where one of the simplest example is 
\begin{eqnarray}
e_{2j-1}=\frac{1}{2}\sigma^{z}_{j},
\hspace{0.3cm}
e_{2j}=\frac{1}{2}\sigma^{x}_{j}\sigma^{x}_{j+1},
\end{eqnarray}
which are the interactions of the transverse Ising chain \cite{62Katsura}-\cite{98Minami}.
The transverse Ising chains 
with random interactions and fields 
have been investigated in \cite{68McCoyWu1}-\cite{11Derzhko}. 
Here we have to note that 
two-dimensional Ising models are equivalent 
to one-dimensional quantum chains \cite{71Suzuki} 
including random cases \cite{14Minami}. 
We can find, for example, 
the cluster models with random next-nearest-neighbor interactions 
are integrable. 
They cannot be diagonalized, 
even in the case of uniform interactions, 
through the standard Jordan-Wigner transformation, 
and the algebraic generalization (\ref{transmain}) is needed 
to diagonalize them \cite{20Yanagihara}. 

The author would like to thank Pascal Baseilhac 
for his valuable comments. 
This work was supported by JSPS KAKENHI Grant No. JP19K03668.

\pagestyle{empty}

\begin{table}
\caption{
Examples of operators 
from which series of operators that satisfy (\ref{condn}) 
can be obtained. 
Here, for example, 
i) 
$XX^{n-1}$ denotes $X_{j}X_{j+1}^{n-1}$ 
ii) 
$X\cdots X$ denotes $1_{j}$ or $X_{j}$ or $X_{j}X_{j+1}$ or 
$\prod_{k=j}^{j+m}X_{k}$ $(m\geq 2)$, 
iii)  
$X\cdots X\cdots X$ denotes $X_{j}$ or $X_{j}X_{j+1}$ or 
$\prod_{k=j}^{j+m}X_{k}$ $(m\geq 2)$. 
When we consider  
$XX^{n-1}$ and $ZZ$, 
let
$\eta^{(1)}_{2j-1}=X_{j}X_{j+1}^{n-1}$,   
$\eta^{(1)}_{2j}=Z_{j+1}Z_{j+2}$, 
and 
$\eta^{(2)}_{2j-1}=X_{j+1}X_{j+2}^{n-1}$,   
$\eta^{(2)}_{2j}=Z_{j+2}Z_{j+3}$. 
Then the series of operators $\{\eta^{(1)}_{j}\}$ satisfy (\ref{condn}), 
and $\{\eta^{(2)}_{j}\}$ satisfy (\ref{condn}), 
and $\eta^{(1)}_{j}$ and $\eta^{(2)}_{k}$ commute with each other 
for all $j$ and $k$. 
}
\begin{tabular}{cc}
\\
\hline
$XX^{n-1}$ & $Z$ 
\\
$XX^{n-1}$ & $ZZ$ 
\\
$XX^{n-1}$ & $Z\cdots Z\cdots Z$ 
\\
$XX^{n-1}$ & $X\cdots XZ\cdots Z\cdots ZX\cdots X$ 
\\
$XX$ & $X\cdots XZZ^{n-1}X\cdots X$ 
\\
$X\cdots X\cdots X$ & $X\cdots XZZ^{n-1}X\cdots X$ 
\\
\\
$YY^{n-1}$ & $Y\cdots YZ\cdots Z\cdots ZY\cdots Y$ 
\\
$Y\cdots Y\cdots Y$ & $Y\cdots YZZ^{n-1}Y\cdots Y$ 
\\
\\
$X1X^{n-1}$ & $Z$ 
\\
$X1X^{n-1}$ & $XZX$ 
\\
$X1\cdots 1X^{n-1}$ & $Z$ 
\\
$X1\cdots 1X^{n-1}$ & $X\cdots XZX\cdots X$ 
\\
\\
$Z1Z^{n-1}$ & $XZX$ 
\\
$Z\underbrace{1\cdots 1}_{l}Z^{n-1}$ & $X\underbrace{Z\cdots Z}_{l}X$ 
\\
\\
$Z1Z$ & $XZX^{n-1}$ 
\\
$Z\underbrace{1\cdots 1}_{l}Z$ & $X\underbrace{Z\cdots Z}_{l}X^{n-1}$ 
\\
\\
$XXZXX^{n-1}$ & $XZX$ 
\\
$XX^{n-1}ZXX^{n-1}$ & $XZX^{n-1}$ 
\\
$XXZ^{n-1}XX^{n-1}$ & $XZ^{n-1}X$ 
\\
$XXZX^{n-1}X^{n-1}$ & $XZ^{n-1}X^{n-1}$ 
\\
\\
$XZZ^{n-1}X^{n-1}$ & $XZX$ 
\\
$XZ^{n-1}ZX^{n-1}$ & $XZ^{n-1}X$ 
\\
$XZ^{n-1}Z^{n-1}X^{n-1}$ & $XZX^{n-1}$ 
\\
$XZZX^{n-1}$ & $XZ^{n-1}X^{n-1}$ 
\\
\\
$XZZZX^{n-1}$ & $XZ^{n-1}X^{n-1}$ 
\\
$XZ^{n-1}ZZ^{n-1}X^{n-1}$ & $XZX^{n-1}$ 
\\
\\
$XZZZX^{n-1}$ & $XZ^{n-1}Z^{n-1}X^{n-1}$ 
\\
$XZZ^{n-1}ZX$ & $XZZ^{n-1}X^{n-1}$ 
\\
\\
$XZ^{n-1}X$ & $YY^{n-1}$ 
\\
$YZ^{n-1}Y$ & $ZZ^{n-1}$ 
\\
$XZ^{n-1}ZX^{n-1}$ & $YZY$ 
\\
\hline
\end{tabular}
\end{table}

\end{document}